# STATISTICAL DECISION THEORY RESPECTING STOCHASTIC DOMINANCE

Charles F. Manski
Department of Economics and Institute for Policy Research, Northwestern University

and

Aleksey Tetenov
Geneva School of Economics and Management, University of Geneva

August 2023

Abstract

The statistical decision theory pioneered by Wald (1950) has used state-dependent mean loss (risk) to measure the performance of statistical decision functions across potential samples. We think it evident that evaluation of performance should respect stochastic dominance, but we do not see a compelling reason to focus exclusively on mean loss. We think it instructive to also measure performance by other functionals that respect stochastic dominance, such as quantiles of the distribution of loss. This paper develops general principles and illustrative applications for statistical decision theory respecting stochastic dominance. We modify the Wald definition of admissibility to an analogous concept of stochastic dominance (SD) admissibility, which uses stochastic dominance rather than mean sampling performance to compare alternative decision rules. We study SD admissibility in two relatively simple classes of decision problems that arise in treatment choice. We reevaluate the relationship between the MLE, James-Stein, and James-Stein positive part estimators from the perspective of SD admissibility. We consider alternative criteria for choice among SD-admissible rules. We juxtapose traditional criteria based on risk, regret, or Bayes risk with analogous ones based on quantiles of state-dependent sampling distributions or the Bayes distribution of loss.

1. Introduction

Wald (1950) considered the broad problem of using sample data to make decisions under uncertainty. He posed the task as choice of a statistical decision function (a rule, for short), which maps potentially available data into a choice among the feasible actions. He recommended ex ante evaluation of statistical decision functions as procedures, chosen prior to realization of the data, specifying how a decision maker would use whatever data may be realized. Expressing the objective as minimization of loss, he proposed that the decision maker evaluate a rule by its mean performance across potential samples, which he termed risk.

In the presence of uncertainty about the loss function and the sampling process yielding the data, Wald prescribed a three-step decision process. The first stage specifies the state space (parameter space), which indexes the loss functions and sampling distributions that the decision maker deems possible. The second stage eliminates inadmissible rules. A rule is inadmissible (weakly dominated) if there exists another one that yields at least as good mean sampling performance in every possible state of nature and strictly better mean performance in some state. The third stage uses some criterion to choose an admissible rule. Wald studied the minimax criterion when the decision maker places no subjective probability distribution on the state space and minimization of Bayes risk (the subjective mean of risk across states) when such a distribution is present.

In many respects, the Wald framework has breathtaking generality. It enables comparison of all statistical decision functions whose risk is well-defined in each possible state. It applies whatever the sampling process and sample size may be. It applies whatever information the decision maker may have about the loss function and the sampling process. The state space may be finite dimensional (parametric) or larger (nonparametric). The true state of nature may be point or partially identified.

A striking exception to the generality of the Wald framework is its use of mean loss to measure the probabilistic performance of alternative rules. Risk is state-dependent mean loss across potential samples and Bayes risk is overall mean loss across samples and states when a subjective distribution is placed on the state space. The literature on statistical decision theory has followed Wald in measuring sampling and overall performance by risk and Bayes risk. See, for example, the texts of Ferguson (1967) and Berger (1985).

We cannot be sure why statistical decision theory has exclusively used mean loss to measure the performance of statistical decision functions, but we can conjecture. One reason may have been the predisposition of statisticians in the mid-twentieth century to use the mean to express the central tendency of probability distributions rather than the median or other location parameters; see Huber (1981) for an interesting discussion. Another reason may have been the influence of the von Neumann and Morgenstern (1944) and Savage (1954) axiomatic derivations of maximization of expected utility, which have often been interpreted as providing rationales to favor this decision criterion over others. Yet subsequent developments in axiomatic decision theory have called into question whether the axioms that yield expected utility maximization are as compelling as they once seemed. See, for example, Binmore (2009).

Considering the matter afresh, we think it evident that evaluation of the probabilistic performance of statistical decision functions should respect stochastic dominance. However, we do not see a compelling reason to focus exclusively on mean loss. We think it instructive to measure probabilistic performance by various functionals that respect stochastic dominance. These include the means of increasing functions of loss and quantiles of the distribution of loss.

This paper develops general principles and illustrative applications for statistical decision theory respecting stochastic dominance. The general principles are introduced in Section 2. We modify the Wald definition of admissibility to an analogous concept of stochastic dominance (SD) admissibility, which uses stochastic dominance rather than mean sampling performance to compare alternative statistical decision functions. We cite representation theorems that characterize stochastic dominance in terms of inequalities ordering the means of increasing functions and the quantiles of two probability distributions. These theorems yield alternative characterizations of SD-admissibility.

Sections 3 and 4 apply the general principles to particular classes of decision problems. Section 3 considers the special case of state-dependent binary loss, where the loss function takes only two values in each state. We show that, when the loss function has this form, state-dependent error probabilities are a sufficient statistic for sampling performance and SD admissibility is equivalent to mean admissibility.

An important application occurs in decision problems where a planner uses sample data to inform choice of one of two treatments to assign to a population of persons. It has been common in medical and other settings to use experimental or observational data on treatment response to test the superiority of one treatment relative to the other and to use the test result to make a treatment choice. In this setting, every rule assigning all members of the population to a single treatment is characterizable as performance of a hypothesis test. We show that the use of error probabilities to determine the admissibility of test rules differs from its traditional use in the theory of hypothesis testing.

Section 4 studies a class of decision problems in which SD and mean admissibility do not coincide. These are problems in which the set of feasible actions is ordered and the sampling process, which generates real-valued data, satisfies the monotone likelihood ratio property. Analysis of mean admissibility in this setting dates back to Karlin and Rubin (1956). Here we study SD admissibility. Possible applications occur when choosing the dose of a real-valued treatment for a population given real-valued sample data that are informative about dose response.

Section 5 revisits the Stein phenomenon of mean inadmissibility of the MLE estimator of a multivariate normal mean of dimension greater or equal to three when the loss function is the component-wise sum of squared losses. We reevaluate the relationship between the MLE, James-Stein, and James-Stein positive part estimators from the perspective of SD admissibility.

Section 6 considers alternative criteria for choice among SD-admissible actions. We juxtapose traditional criteria based on risk, regret, or Bayes risk with analogous ones based on quantiles of state-dependent sampling distributions or the Bayes distribution of loss. We show how mean and quantile criteria differ when applied to choice of a test rule.

2. General Principles

Section 2.1 reviews the concepts of Wald's statistical decision theory. Section 2.2 generalizes these concepts to make stochastic dominance rather than risk the basic quantity used to evaluate the performance of statistical decision functions. Section 2.3 uses two representation theorems for stochastic dominance to characterize SD-admissibility by classes of inequalities that order the means of increasing functions and the quantiles of loss.

2.1. Concepts of the Wald Theory

Wald's statistical decision theory begins with specification of a state space S, a set of feasible decisions (or actions) D, and a loss function $L(\cdot, \cdot): S \times D \to [0, \infty)$ specifying the loss incurred by each feasible action in each possible state. The ideal objective is to minimize loss in the true state. Given that the true state is unknown, the ideal objective is sure to be achievable only if there exists an action that uniformly minimizes loss in all states of S. Wald's practical objective is to prescribe reasonable decision rules when no such action exists.

The adjective "statistical" describes statistical decision theory because Wald assumes that a state-dependent sampling distribution $Q_s$ generates data whose value, say $\psi$, lies in a known sample space $\Psi$. He supposes that the decision maker observes $\psi$ and knows the vector $(Q_s, s \in S)$ of state-dependent sampling distributions. In this setting, a statistical decision function $\delta(\cdot): \Psi \to D$ is any $\Psi$-measurable function that maps the data into an action. Let $\Delta$ denote the space of feasible rules.

Research in statistical decision theory often finds it useful to consider randomized rules that map the data into a specified probability distribution on D rather than into a specific action. Consideration of randomized rules does not require alteration of the definition of $\delta$. One may define the sample space and the state-dependent sampling distributions to include a white-noise component used to randomly choose an action.

To measure the performance of a candidate rule $\delta$, Wald focuses on the state-dependent mean loss (risk) that it generates across potential samples; that is,

$$(1) \quad R(s, \delta) \equiv E_s\{L[s, \delta(\psi)]\} = \int L[s, \delta(\psi)] dQ_s(\psi).$$

Risk is computable in principle, although computation may be difficult in practice. Supposing that computation of risk is tractable, Wald recommends use of the vector $[R(s, \delta), s \in S]$ of state-dependent risks to measure the performance of $\delta$ across potential samples and to compare $\delta$ with other rules.

To begin, rule $\delta$ is deemed better than rule $\delta'$ if $R(s, \delta) \leq R(s, \delta')$ for all $s \in S$ and $R(s, \delta) < R(s, \delta')$ for some s. If there exists a $\delta$ that is better than $\delta'$ in this sense, then $\delta'$ is said to be inadmissible and should be eliminated from further consideration. A rule that is not inadmissible is called admissible.

Going a bit further, a decision maker can eliminate an admissible rule when there exists a risk-equivalent rule that is retained for consideration. Rules $\delta$ and $\delta'$ are risk-equivalent if $R(s, \delta) = R(s, \delta')$ for all $s \in S$. When multiple admissible rules are risk-equivalent, a decision maker who uses risk to evaluate sampling performance can eliminate all but one of them without consequence.

Having eliminated all inadmissible rules and perhaps some admissible rules within risk-equivalent groups of rules, the decision maker's problem is to choose among the subset of rules that remain, say $\Delta_a$.

It is possible in principle that $\Delta_a$ may be empty, but applications of the Wald theory typically have enough regularity to ensure not only that $\Delta_a$ is non-empty but that every inadmissible decision function is dominated by an admissible one.

Whereas elimination of inadmissible and risk-equivalent admissible rules is uncontroversial, there is no consensus on choice within $\Delta_a$, which requires comparison of rules whose risk vectors are unordered. Wald studied minimization of Bayes risk when the decision maker places a subjective probability distribution, say $\pi$, on the state space. This criterion solves the problem

$$(2) \quad \min_{\delta \in \Delta_a} \int R(s, \delta) d\pi(s).$$

Wald suggested the minimax criterion in the absence of a subjective distribution on S, stating (p. 18) "a minimax solution seems, in general, to be a reasonable solution of the decision problem when an a priori distribution . . . . does not exist or is unknown to the experimenter." This criterion solves

$$(3) \quad \min_{\delta \in \Delta_a} \max_{s \in S} R(s, \delta).$$

Viewing the minimax criterion as unduly conservative, Savage (1951) suggested the minimax-regret criterion in his review essay on Wald (1950). This criterion solves

$$(4) \quad \min_{\delta \in \Delta_a} \max_{s \in S} [R(s, \delta) - \min_{d \in D} L(s, d)].$$

Subsequent research applying statistical decision has generally used criterion (2), (3), or (4).

It often is difficult to determine the set of admissible rules. Given this, researchers applying the Wald theory commonly skip the step of determining admissibility and use a decision criterion to choose among all feasible options, not just those that are admissible. When any of the criteria listed above yields a unique choice, it necessarily is admissible. When a criterion yields a set of equally good choices, the set may include inadmissible options that are strictly dominated only in states that do not affect the value of the optimum. Bayes risk is unaffected by values of risk that occur off the π-support of S. Maximum risk and regret are unaffected by dominance in states that do not determine the maximum.

2.2. Respect for Stochastic Dominance

The new work of this paper begins with the observation that the basic probabilistic quantity underlying statistical decision theory is not risk but rather the state-dependent distribution of loss that a decision function generates across potential samples; that is, $Q_s\{L[s, \delta(\psi)]\}$. The expectation (risk) is but one of many potentially relevant features of this distribution.

State-dependent distributions of loss are computable in principle. Supposing that computation is tractable, we think it natural to generalize the Wald theory by recommending use of the vector $(Q_s\{L[s, \delta(\psi)]\}, s \in S)$ to measure the performance of $\delta$ across potential samples. It is also natural to recommend that evaluation of the performance of alternative statistical decision functions should respect stochastic dominance. This recommendation has many precedents in studies of decision making that are not explicitly concerned with use of sample data. See, for example, Quirk and Saposnik (1962), Hadar and Russell (1969), Hanoch and Levy (1969), and Manski (1988).

Let $P \geq_{sd} P'$ denote that distribution P either equals or stochastically dominates P′, and let $P >_{sd} P'$ denote that P stochastically dominates P′. When considering state-dependent sampling performance, respect for stochastic dominance means that one should prefer rule $\delta$ to $\delta'$ if $Q_s\{L[s, \delta'(\psi)]\} \geq_{sd} Q_s\{L[s, \delta(\psi)]\}$ for all $s \in S$ and $Q_s\{L[s, \delta'(\psi)]\} >_{sd} Q_s\{L[s, \delta(\psi)]\}$ for some s. We will adapt Wald's definition of (mean) admissibility and say that δ′ is stochastic-dominance inadmissible (*SD-inadmissible*) if these conditions hold.

When a decision maker places a subjective distribution on the state space, respect for stochastic dominance means that one should Bayes-SD prefer δ to δ′ if the distribution of loss across samples and states generated by δ′ stochastically dominates that generated by δ. The distribution of loss under δ is

(5)  $\Phi_\pi\{L[s, \delta(\psi)]\} \equiv \int Q_s\{L[s, \delta(\psi)]\} d\pi(s)$.

We will adapt Wald's definition of Bayes risk and say that $\Phi_\pi$ is the Bayes loss distribution.

The fact that $\Phi$ is the mean over S of the state-dependent loss distributions implies a connection between SD-preference and Bayes-SD preference. The following lemma follows immediately from (5):

*Lemma 1*: If rule δ is SD-preferred to rule δ′, then $\Phi_\pi\{L[s, \delta'(\psi)]\} \geq_{sd} \Phi_\pi\{L[s, \delta(\psi)]\}$.  ☐

2.3. Representation Theorems Relating SD-Admissibility to Mean and Quantile Loss

Respect for stochastic dominance does not require the decision maker to use any particular real functional of loss distributions to measure the performance of a decision function. Nevertheless, there exist useful representation theorems that characterize stochastic dominance in terms of two alternative classes of functionals, these being means of increasing functions of loss and quantiles of the distribution of loss.

2.3.1. Means of Increasing Functions of Loss

Let P and P′ denote two probability distributions on the real line. It has long been known that P = P′ if and only if $\int f(y)dP(y) = \int f(y)dP'(y)$ for every integrable increasing function f(·). Several articles studying expected utility maximization when utility is an increasing function of income show that P stochastically dominates P′ if and only if $\int f(y)dP(y) \geq \int f(y)dP'(y)$ for every integrable increasing function $f(\cdot)$ and $\int f(y)dP(y) > \int f(y)dP'(y)$ for some increasing $f(\cdot)$. See Quirk and Saposnik (1962), Hadar and Russell (1969), and Hanoch and Levy (1969). This representation theorem immediately yields a characterization of SD-inadmissibility:

*Lemma 2*: Rule $\delta'$ is SD-inadmissible if and only if there exists a rule $\delta$ such that

(6) $E_s(f\{L[s, \delta'(\psi)]\}) \geq E_s(f\{L[s, \delta(\psi)]\})$ for every increasing $f(\cdot)$ and all $s \in S$,

$E_s(f\{L[s, \delta'(\psi)]\}) > E_s(f\{L[s, \delta(\psi)]\})$ for some increasing $f(\cdot)$ and s.  □

2.3.2. Quantiles of Loss

Let P denote a probability distribution on the real line. For $\lambda \in (0, 1)$, let $V_\lambda(P)$ denotes the $\lambda$-quantile of P. It has long been known that $P = P'$ if and only if $V_\lambda(P) = V_\lambda(P')$ for all $\lambda \in (0, 1)$. Levy and Kroll (1978) show that P′ stochastically dominates P if and only if $V_\lambda(P') \geq V_\lambda(P)$ for all $\lambda \in (0, 1)$ and $V_\lambda(P') > V_\lambda(P)$ for some $\lambda \in (0, 1)$. This representation theorem immediately yields another characterization of SD-inadmissibility.

*Lemma 3*: Rule $\delta'$ is SD-inadmissible if and only if there exists a $\delta$ such that

(7) $V_\lambda(Q_s\{L[s, \delta'(\psi)]\}) \geq V_\lambda(Q_s\{L[s, \delta(\psi)]\})$ for all $\lambda \in (0, 1)$ and $s \in S$,

$V_\lambda(Q_s\{L[s, \delta'(\psi)]\}) > V_\lambda(Q_s\{L[s, \delta(\psi)]\})$ for some $\lambda \in (0, 1)$ and s.  □

## 3. State-Dependent Binary Loss

SD-inadmissibility and mean inadmissibility are equivalent to one another when the loss function has a special form that occurs in some important applications. Partition the set D of feasible actions into two mutually exclusive subsets, say $D_a$ and $D_b$. Let $L_{sa}$ and $L_{sb}$ be a specified pair of state-dependent real numbers. The loss function has the state-dependent binary form if

(8)  $L[s, \delta(\psi)] = L_{sa} \cdot 1[\delta(\psi) \in D_a] + L_{sb} \cdot 1[\delta(\psi) \in D_b].$

Section 3.1 develops the basic finding. Section 3.2 applies it to choice between two treatments.

### 3.1. Using Error Probabilities to Characterize SD and Mean Admissibility

When loss has form (8), its state-dependent sampling distribution is determined by the state-dependent choice probabilities $Q_s[\delta(\psi) \in D_a]$ and $Q_s[\delta(\psi) \in D_b]$, which sum to one. Rule $\delta$ has a state-dependent distribution placing mass $Q_s[\delta(\psi) \in D_a]$ at $L_{sa}$ and mass $Q_s[\delta(\psi) \in D_b]$ at $L_{sb}$. Hence, rule $\delta$ is SD-preferred to $\delta'$ if and only if $\delta$ places weakly more mass than $\delta'$ at min $(L_{sa}, L_{sb})$ in every state and strictly more mass in some state where $L_{sa} \neq L_{sb}$.

A succinct way to express SD-preference is to define the state-dependent probability $\rho_s(\delta)$ that $\delta$ yields an error, choosing the action with larger loss rather than the one with smaller loss. An error is logically impossible when $L_{sa} = L_{sb}$, so $\rho_s(\delta) = 0$ in these states. In states with $L_{sa} \neq L_{sb}$,

(9) $\quad \rho_s(\delta) \quad = Q_s[\delta(\psi) \in D_a]$ if $L_{sa} > L_{sb}$,

$\quad\quad\quad\quad\quad = Q_s[\delta(\psi) \in D_b]$ if $L_{sb} > L_{sa}$.

With this definition of error probabilities, we obtain a simple characterization of SD-inadmissibility.

*Lemma 4*: Let the loss function have form (8). Then rule $\delta'$ is SD-inadmissible if and only if there exists another rule $\delta$ such that $\rho_s(\delta) \leq \rho_s(\delta')$ for all $s \in S$ and $\rho_s(\delta) < \rho_s(\delta')$ for some s.  □

Error probabilities also characterize mean admissibility. Given a loss function of form (8), mean loss in state s (risk) is

(10) $\int L[s, \delta(\psi)]dQ_s(\psi) \;=\; L_{sa} \cdot Q_s[\delta(\psi) \in D_a] + L_{sb} \cdot Q_s[\delta(\psi) \in D_b]$

$\quad\quad\quad\quad\quad\quad\quad\quad = \min(L_{sa}, L_{sb})[1 - \rho_s(\delta)] + \max(L_{sa}, L_{sb}) \cdot \rho_s(\delta)$.

This yields a parallel characterization of mean inadmissibility.

*Lemma 5*: Let the loss function have form (8). Then rule $\delta'$ is mean inadmissible if and only if there exists another rule $\delta$ such that $\rho_s(\delta) \leq \rho_s(\delta')$ for all $s \in S$ and $\rho_s(\delta) < \rho_s(\delta')$ for some s.  □

3.2. Choice between Two Treatments

An important class of applications of statistical decision theory consider use of sample data on treatment response to inform a planner who must choose treatments for a population. Past work by Manski

(2004, 2005, 2021), Manski and Tetenov (2007), Hirano and Porter (2009), Stoye (2009, 2012), Tetenov (2012), Manski and Tetenov (2016), and Kitagawa and Tetenov (2018) has used the Wald framework to study this decision problem. A statistical decision function uses the data to choose a treatment allocation, so such a function has been called a *statistical treatment rule* (STR). The planner's objective has been expressed as maximization of a social welfare function that sums treatment outcomes across the population. The mean sampling performance of an STR has been called *expected welfare*. Maximization of social welfare is equivalent to minimization of loss. Expected welfare is negative risk.

We consider here the relatively simple case in which the planner must assign one of two treatments to each member of a treatment population, denoted J. The feasible treatments are $T = \{a, b\}$. Each $j \in J$ has a response function $u_j(\cdot): T \to Y$ mapping treatments $t \in T$ into real-valued individual welfare outcomes $u_j(t)$. Treatment is individualistic; that is, a person's outcome may depend on the treatment he is assigned but not on the treatments assigned to others. The population is a probability space $(J, \Omega, P)$, and the probability distribution $P[u(\cdot)]$ of the random function $u(\cdot): T \to R$ describes treatment response across the population. The population is large in the sense that J is uncountable and $P(j) = 0$, $j \in J$.

While treatment response may be heterogeneous, we suppose here that the members of the population are observationally identical to the planner. That is, the planner does not observe person-specific covariates that would enable systematic differentiation of treatment of different persons. In principle, the planner can randomly allocate persons to the two treatments with specified allocation probabilities. The notation introduced below allows for this possibility. However, when applying the findings of Section 3.1, we will consider only *test rules*, which assign all members of the population to one treatment or the other.

3.2.1. The Mean Sampling Performance of STRs

A statistical treatment rule maps sample data into a treatment allocation. Let $\Delta$ denote the space of functions that map $T \times \Psi$ into the unit interval and that satisfy the adding-up conditions: $\delta \in \Delta \Rightarrow \delta(a, \psi)$

+ δ(b, ψ) = 1, ∀ ψ ∈ Ψ. Each function δ ∈ Δ defines a statistical treatment rule, δ(a, ψ) and δ(b, ψ) being the fractions of the population assigned to treatments a and b when the data are ψ. Observe that this definition of an STR does not specify which persons receive each treatment, only the assignment shares. Designation of the particular persons receiving each treatment is immaterial because assignment is random, the population is large, and the planner has an additive welfare function. As δ(a, ψ) + δ(b, ψ) = 1, we use the shorthand δ(ψ) to denote the fraction assigned to treatment b. The fraction assigned to treatment a is 1 − δ(ψ).

The planner wants to maximize population welfare, which adds welfare outcomes across persons. Given data ψ, the population welfare (negative loss) realized if the planner were to choose rule δ is

(11) $\quad U(\delta, P, \psi) \equiv E[u(a)] \cdot [1 - \delta(\psi)] + E[u(b)] \cdot \delta(\psi) \equiv \alpha \cdot [1 - \delta(\psi)] + \beta \cdot \delta(\psi),$

where $\alpha \equiv E[u(a)] \equiv \int_J u_j(a) dP(j)$ and $\beta \equiv E[u(b)] \equiv \int_J u_j(b) dP(j)$ are assumed to be finite. Inspection of (11) shows that, whatever value ψ may take, it is optimal to set $\delta(\psi) = 0$ if $\alpha \geq \beta$ and $\delta(\psi) = 1$ if $\alpha \leq \beta$.

The problem of interest is treatment choice when knowledge of P and Q does not suffice to determine the ordering of α and β. Hence, the planner does not know the optimal treatment. Let {($P_s$, $Q_s$), s ∈ S} be the set of (P, Q) pairs that the planner deems possible. The planner does not know the optimal treatment if S contains at least one state such that $\alpha_s > \beta_s$ and another such that $\alpha_s < \beta_s$. We assume this throughout.

Considered as a function of ψ, $U(\delta, P_s, \psi)$ is a random variable with state-dependent sampling distribution $Q_s[U(\delta, P_s, \psi)]$. Following Wald's view of statistical decision functions as procedures, we use the vector $\{Q_s[U(\delta, P_s, \psi)], s \in S\}$ of state-dependent welfare distributions to evaluate rule δ. In principle this vector is computable, whatever the state space and sampling process may be. Hence, in principle, a planner can compare the vectors of state-dependent welfare distributions yielded by different STRs and base treatment choice on this comparison.

Respect for stochastic dominance means that the planner prefers rule $\delta$ to an alternative rule $\delta'$ if $Q_s[U(\delta, P_s, \psi)] \geq_{sd} Q_s[U(\delta', P_s, \psi)]$ for all $s \in S$ and $Q_s[U(\delta, P_s, \psi)] >_{sd} Q_s[U(\delta', P_s, \psi)]$ for some s. The expected welfare (negative risk) of rule $\delta$ in state s, denoted $W(\delta, P_s, Q_s)$, is

(12) $\quad W(\delta, P_s, Q_s) = \alpha_s \cdot \{1 - E_s[\delta(\psi)]\} + \beta_s \cdot E_s[\delta(\psi)],$

where $E_s[\delta(\psi)] \equiv \int_\Psi \delta(\psi) dQ_s(\psi)$ is the mean (across potential samples) fraction of persons who are assigned to treatment b.

3.2.2. SD and Mean Admissibility of Test Rules

An important class of STRs are the *uniformly singleton* rules. Given a treatment set of any size, a rule is uniformly singleton if, for every possible data realization, it assigns the entire population to one treatment. The treatment to which the entire population is assigned may vary with the data realization.

Uniformly singleton rules are particularly simple when there are two treatments. In this case, a rule is uniformly singleton rule if, for each $\psi \in \Psi$, $\delta(\psi) = 1$ or $\delta(\psi) = 0$. The class of uniformly singleton STRs is the same as the class of rules that use the outcome of a hypothesis test to choose between the treatments.

Construction of a *test rule* begins by partitioning the state space into disjoint subsets $S_a$ and $S_b$, where $S_a$ contains all states in which treatment a is uniquely optimal and $S_b$ contains all states in which b is uniquely optimal. Thus, $\alpha_s > \beta_s \Rightarrow s \in S_a$, $\alpha_s < \beta_s \Rightarrow s \in S_b$, and the states with $\alpha_s = \beta_s$ are somehow split between the two sets. Let s* denote the unknown true state. The two hypotheses are [s* $\in S_a$] and [s* $\in S_b$].

A test rule $\delta$ partitions the sample space $\Psi$ into disjoint acceptance regions $\Psi_{\delta a}$ and $\Psi_{\delta b}$. When the data $\psi$ lie in $\Psi_{\delta a}$, the rule accepts hypothesis $[s^* \in S_a]$ by setting $\delta(\psi) = 0$. When $\psi$ lies in $\Psi_{\delta b}$, the rule accepts $[s^* \in S_b]$ by setting $\delta(\psi) = 1$. We use the word "accepts" rather than the traditional term "does not reject" because treatment choice is an affirmative action.

The above shows that test rules are uniformly singleton. The converse holds as well. If $\delta$ is uniformly singleton, one can collect all of the data values for which the rule assigns everyone to treatment a, call this subset of the sample space the acceptance region $\Psi_{\delta a}$, and do likewise for $\Psi_{\delta b}$. In what follows, we use the term *test* rule rather than *uniformly singleton* rule.

Test rules have the state-dependent binary form (8), with $L_{sa} = -\alpha_s$, $L_{sb} = -\beta_s$, $D_a = 0$ and $D_b = 1$. An error is impossible when $\alpha_s = \beta_s$, so $\rho_s(\delta) = 0$ in these states. In states with $\alpha_s \neq \beta_s$, the error probability is

$$
\begin{aligned}
(13) \quad \rho_s(\delta) &= Q_s(\psi \in \Psi_{\delta a}) \text{ if } \alpha_s < \beta_s, \\
&= Q_s(\psi \in \Psi_{\delta b}) \text{ if } \alpha_s > \beta_s.
\end{aligned}
$$

Lemmas 4 and 5 shows that a test rule $\delta$ is both SD and mean inadmissible if there exists another test rule $\delta'$ such that $\rho_s(\delta) \leq \rho_s(\delta')$ for all $s \in S$ and $\rho_s(\delta) < \rho_s(\delta')$ for some s.

A special but important class of hypothesis tests juxtaposes two simple hypotheses. Then the Neyman-Pearson Lemma shows that, among all tests with a specified probability of a Type I error, the likelihood-ratio test minimizes the probability of a Type II error, and vice versa. In the context of treatment choice, having two simple hypotheses means that S contains two states, with treatment a better in one state and b better in the other. Then the Neyman-Pearson Lemma implies that a planner considering use of a test rule need not look beyond the class of likelihood-ratio tests. Applying Lemmas 4 and 5 to likelihood ratio tests yields this result, which makes explicit the form of error probabilities for likelihood-ratio tests.

*Lemma 6*: Let S = {0, 1}, with $\alpha_0 > \beta_0$ and $\alpha_1 < \beta_1$. Let the data have distinct state-dependent sampling distributions $Q_0$ and $Q_1$ with either Lebesgue density or probability mass functions $q_0(\cdot)$ and $q_1(\cdot)$. Let $\delta$ be a test rule. For $\eta \geq 0$, let $\delta(\eta)$ be the likelihood-ratio rule with threshold $\eta$; thus, $\Psi_{\delta(\eta)a} = [\psi \in \Psi: q_1(\psi) \leq \eta q_0(\psi)]$ and $\Psi_{\delta(\eta)b} = [\psi \in \Psi: q_1(\psi) > \eta q_0(\psi)]$. Rule $\delta$ is both SD and mean inadmissible if there exists an $\eta \geq 0$ such that $\rho_0(\delta) \geq Q_0[q_1(\psi) > \eta q_0(\psi)]$, $\rho_1(\delta) \geq Q_1[q_1(\psi) \leq \eta q_0(\psi)]\}$, and at least one inequality is strict.

□

*Proof*: Rule $\delta(\eta)$ has error probabilities $\rho_0[\delta(\eta)] = Q_0[q_1(\psi) > \eta q_0(\psi)]$ and $\rho_1[\delta(\eta)] = Q_1[q_1(\psi) \leq \eta q_0(\psi)]$. Hence, the result is an immediate application of the proposition.

Q. E. D.

A fundamental feature of the above analysis is that all error probabilities symmetrically determine the result. In contrast, the classical theory of hypothesis testing differentiates between null and alternative hypotheses, and correspondingly between Type I and Type II errors. It restricts attention to tests that yield a predetermined probability of a Type I error and seeks a test of this type that yields an adequately small probability of a Type II error.

For example, a document of the U. S. Food and Drug Administration providing guidance for the design of randomized clinical trials (RCTs) evaluating new medical devices states that the probability of a Type I error is conventionally set to 0.05 and that the probability of a Type II error depends on the claim for the device but should not exceed 0.20 (U. S. Food and Drug Administration, 2014). The International Conference on Harmonisation (ICH) has provided similar guidance for the design of RCTs evaluating pharmaceuticals. The ICH document states the following (International Conference on Harmonization, 1999, p. 1923):

> "Conventionally the probability of type I error is set at 5% or less or as dictated by any adjustments made necessary for multiplicity considerations; the precise choice may be influenced by the prior plausibility of the hypothesis under test and the desired impact of the results. The probability of type II error is conventionally set at 10% to 20%; it is in the sponsor's interest to keep this figure as low as feasible especially in the case of trials that are difficult or impossible to repeat. Alternative values to the conventional levels of type I and type II error may be acceptable or even preferable in some cases."

Such asymmetric treatment of the two hypotheses is illogical from the perspective of statistical decision theory.

4. Ordered Actions and Continuous Real Data Satisfying the Monotone-Likelihood Ratio Property

We now study a class of decision problems in which SD and mean admissibility do not coincide. These are problems in which the set of feasible actions is ordered and the sampling process generating the data satisfies the monotone likelihood ratio property. Analysis of mean admissibility in this setting dates back to Karlin and Rubin (1956), with continuation by Manski and Tetenov (2007). Here we study SD admissibility. Section 4.1 develops the basic finding. Section 4.2 applies it to treatment choice.

4.1. Basic Finding

Proposition 7 shows that the fractional monotone treatment rules form an essentially complete class with respect to stochastic dominance when the data satisfy the maintained assumptions. A *fractional monotone* rule is one in which $\delta(\psi)$ is weakly increasing in $\psi$. *Essential completeness* means that any randomized decision rule $\delta(\psi, \upsilon)$ can be replaced by a fractional monotone rule $\delta'(\psi)$ that weakly stochastically dominates $\delta(\psi, \upsilon)$ in each state s. The planner then does not need to consider any other types of STRs. Manski and Tetenov (2007, Proposition 1) show that fractional monotone rules form an essentially complete class when the planner wants to maximize the expectation $E_s[f(U(\delta, P_s, \psi, \upsilon))]$ of a concave-monotone function $f(\cdot)$ of the population welfare and $\psi$ is binomial. Here we establish a more general result that a planner with any decision criterion that respects stochastic dominance can restrict attention to fractional monotone rules.

*Proposition 7:* Let $u(a, s)$ be the payoff function from action $a \in [a_l, a_h] \subset R$ in state $s \in S$. Assume that $u(a, s)$ is weakly monotonic in a for each s. For $s \in S$, let the data $\psi \in R$ have a continuous distribution $Q_s(\psi)$ and density $q_s(\psi)$ with respect to Lebesgue measure. Let $\upsilon \sim \text{Uniform}[0, 1]$ be a randomization variable independent of $\psi$. Assume that there exists a state $s_0$ for which $u(a, s_0)$ is constant in a.

Let $Q_s(\psi)$ possess the *monotone likelihood ratio property* for all pairs $(s, s_0)$ such that $u(a, s)$ is not constant in a. That is,

(14a)   if $u(a, s)$ is non-increasing in a, then $q_s(\psi)/q_{s_0}(\psi) \geq q_s(\psi')/q_{s_0}(\psi')$ for all $\psi < \psi'$,

(14b)   if $u(a, s)$ is non-decreasing in a, then $q_s(\psi)/q_{s_0}(\psi) \leq q_s(\psi')/q_{s_0}(\psi')$ for all $\psi < \psi'$.

Then for any randomized strategy $\delta(\psi, \upsilon): \Psi \times [0, 1] \to [a_l, a_h]$, there exists a monotone non-randomized strategy $\delta'(\psi): \Psi \to [a_l, a_h]$ whose distribution of payoffs $Q_s[u(\delta'(\psi),s)]$ weakly stochastically dominates the distribution of payoffs $Q_s[u(\delta(\psi, \upsilon),s)]$ of $\delta$ in each state s. $\delta'$ could be constructed by monotonically rearranging the values taken by $\delta(\psi, \upsilon)$ in state $s_0$:

(15)    $\delta'(\psi) \equiv G^{-1}_{\delta,s_0}(F_0(\psi))$,

where $G^{-1}_{\delta,s_0}(\cdot)$ is the quantile function of the distribution $Q_{s_0}[\delta(\psi, \upsilon)]$ of the action $\delta(\psi, \upsilon)$ in state $s_0$ and $F_0(t) \equiv Q_{s_0}(\psi \leq t)$ is the c.d.f. of $\psi$ in state $s_0$. □

*Proof*: First we show that the non-randomized strategy $\delta'(\psi)$ defined in (15) re-arranges the values of $\delta(\psi, \upsilon)$ to be increasing in $\psi$ and has the same probability distribution of actions (and hence payoffs) as $\delta(\psi, \upsilon)$ in state $s_0$. Then we show that $Q_s[u(\delta'(\psi),s)]$ weakly stochastically dominates $Q_s[u(\delta(\psi, \upsilon),s)]$ in all other states of nature.

Given that $\psi$ has a continuous distribution, random variable $F_0(\psi)$ has a Uniform(0, 1) distribution in state $s_0$. Hence, random variable $\delta'(\psi) = G^{-1}_{\delta,s_0}(F_0(\psi))$ has c.d.f. $G_{\delta,s_0}$ in state $s_0$.[1] Given that both $G^{-1}_{\delta,s_0}(\cdot)$ and $F_0(\cdot)$ are non-decreasing, $\delta'(\psi)$ is also non-decreasing in $\psi$. Given that $F_0$ is continuous and $G^{-1}_{\delta,s_0}$ is left-continuous, $\delta'(\psi)$ is also left-continuous.

In states where $u(a, s)$ is constant in $a$, the distributions of payoffs are identical for all strategies. Hence, weak stochastic dominance holds. Now suppose that state $s$ satisfies (14a), so $u(a, s)$ is non-increasing in $a$. (The proof is analogous for states in which $u(a, s)$ is non-decreasing in $a$.)

We want to show that the distribution of $\delta'(\psi)$ is weakly stochastically dominated by the distribution of $\delta(\psi, \upsilon)$. Denote the c.d.f. of action $\delta(\psi, \upsilon)$ in state $s$ by

$$G_{\delta,s}(t) \equiv Q_s[u(\delta(\psi, \upsilon),s)] = \int q_s(\psi) \int 1[\delta(\psi, \upsilon) \leq t] d\upsilon d\psi.$$

---

[1] Let Q and F be the quantile and distribution functions of a random variable. It is the case that, for all u in (0, 1) and all real t (see, for example, Pfeiffer, 1990, p.266) $Q(u) \leq t \Leftrightarrow u \leq F(t)$. If u is itself random with distribution P, it follows that $P[Q(u) \leq t] = P[u \leq F(t)]$. If u is uniform, $P[u \leq F(t)] = F(t)$.

Given any t ∈ [$a_l$, $a_h$], consider the indicator functions $1[\delta(\psi, \upsilon) \leq t]$ and $1[\delta'(\psi) \leq t]$. These indicator functions generate rejection regions for classical hypothesis tests with null hypothesis $s_0$ and alternative hypothesis s. A randomized test with rejection region $\Omega \equiv \{(\psi, \upsilon): \delta(\psi, \upsilon) \leq t\}$ has power function $G_{\delta,s}(t)$ as a function of s. Similarly, $G_{\delta',s}(t) \equiv \int q_s(\psi) \cdot 1[\delta'(\psi) \leq t]d\psi$ is the power function (as a function of s) of a non-randomized test with rejection region $\Omega' \equiv \{\psi: \delta'(\psi) \leq t\}$. We have shown above that the two tests have equal power in state $s_0$: $G_{\delta',s_0}(t) = G_{\delta,s_0}(t)$. Given that $\delta'(\psi)$ is non-decreasing in $\psi$, $1[\delta'(\psi) \leq t]$ is non-increasing in $\psi$ and there exists $\psi_t$ such that

$1[\delta'(\psi) \leq t] = 1$ for all $\psi < \psi_t$,

$1[\delta'(\psi) \leq t] = 0$ for all $\psi > \psi_t$.

Given that state s satisfies (14a), the test with rejection region $\Omega' = (\psi: \psi \leq \psi_t)$ is a likelihood-ratio test. The tests with rejection regions $\Omega$ and $\Omega'$ have the same size. If follows from the Neyman-Pearson lemma that test $\Omega'$ must be at least as powerful as $\Omega$ in state s.[2] That is, $G_{\delta',s}(t) \geq G_{\delta,s}(t)$.

We can thus establish that $G_{\delta',s}(t) \geq G_{\delta,s}(t)$ for all t. Hence, the distribution of $\delta(\psi, \upsilon)$ weakly stochastically dominates the distribution of $\delta'(\psi)$. Given that u(a, s) is a weakly decreasing function of a, $Q_s[u(\delta'(\psi),s)]$ weakly stochastically dominates $Q_s[u(\delta(\psi, \upsilon),s)]$.

Q. E. D.

4.2. Applications to Treatment Choice

4.2.1. Choice between Two Treatments

---

[2] For a version that covers randomized tests see, for example, Lehmann and Romano (2008), Theorem 3.2.1.

Proposition 7 applies to the treatment-choice problem of Section 3.2, with action a ∈ [0, 1] denoting the fraction of the population assigned to treatment b. Payoff function (12) is decreasing in a when $\beta_s - \alpha_s < 0$, increasing in a when $\beta_s - \alpha_s > 0$, and constant when $\beta_s - \alpha_s = 0$. Hence, the payoff function satisfies the assumptions of the proposition. Suppose that $Q_s(\psi)$ is continuous and possesses the monotone likelihood ratio property in $(\beta - \alpha)$. Then the proposition shows that the class of fractional monotone STRs is essentially complete under any decision criterion that respects stochastic dominance.

4.2.2. Choice of Treatment Dose

Let action a be a dose level for a real-valued treatment; for example, it may be the dose of a medical drug treatment. Suppose that administering a higher dose is beneficial but costly. In particular, let the payoff function have the linear form $u(a, s) = b(s)a - ca$, which gives the benefit minus cost of different dose levels. Here c is the known marginal cost of increasing the dose and $b(s)$ is the unknown marginal dose response. Then $u(a, s)$ is increasing if $b(s) > c$ and decreasing if $b(s) < c$. Welfare does not vary with the dose level when $b(s) = c$. Thus, $u(\cdot, \cdot)$ satisfies the assumptions of Proposition 7.

Suppose that one obtains real-valued data $\psi$ drawn from a continuous distribution and that $\psi$ provides an informative but imperfect signal about $b(s)$; for example, $\psi$ may be the result of an informative but imperfect diagnostic test. It is relatively easy to imagine signal generation processes in which $\psi$ has the MLR property. For example, it may be that $\psi$ equals $b(s)$ plus a white-noise error. Then Proposition 7 implies that treatment rule should be monotone in the realization of $\psi$ and should be non-randomized.

5. The Stein phenomenon from the perspective of stochastic dominance

This section studies another class of decision problems in which SD admissibility and mean admissibility do not coincide. We reappraise the Stein phenomenon (Stein, 1956) from the perspective of SD admissibility. We consider the simplest setting in which the Stein phenomenon occurs. We are interested in estimating a three-dimensional parameter $\theta = (\theta_1 \; \theta_2 \; \theta_3)' \in \mathbb{R}^3$. We observe a three-dimensional outcome $x = (x_1 \; x_2 \; x_3)'$ whose sampling distribution is multivariate normal with mean vector $\theta$ and a known identity variance matrix, $x \sim \mathcal{N}(\theta, I_3)$. The statistical decision rule maps the outcome into a three-dimensional estimator $\delta(x) = \big(\delta_1(x) \; \delta_2(x) \; \delta_3(x)\big)'$ and the loss function is the component-wise sum of squared losses $L(\theta, \delta) \equiv \|\delta - \theta\|^2 = \sum_{i=1}^{3}(\delta_i - \theta_i)^2$.

The MLE estimator in this problem is $\delta_{MLE}(x) \equiv x$. Stein (1956) has shown that the MLE estimator is mean-inadmissible. It is mean-dominated, for example, by the James-Stein (1961) estimator $\delta_{JS} \equiv \left(1 - \frac{1}{\|x\|^2}\right)x$. For any $\theta \in \mathbb{R}^3$, $E_\theta L(\theta, \delta_{JS}) < E_\theta L(\theta, \delta_{MLE})$. Other estimators mean-dominating the MLE estimator have been subsequently proposed. Among those we will consider the James-Stein positive part estimator $\delta_{JSPP} \equiv max\left(0, 1 - \frac{1}{\|x\|^2}\right)x$ from Efron and Morris (1973).

Does the loss distribution of the James-Stein estimator stochastically dominate the loss distribution of the MLE estimator? To show that the statement is false it suffices to find one value of $\theta$ for which $Q_s\{L[\theta, \delta_{JS}]\} \not\succeq_{sd} Q_s\{L[\theta, \delta_{MLE}]\}$. While we cannot offer a mathematical proof of this statement, we have convincing evidence from simulations that stochastic dominance does not hold for $\theta = [0 \; 0 \; 0]'$.

Figure 1 plots the cumulative distribution of loss for the MLE, James-Stein, and James-Stein positive part estimators at $\theta = [0 \; 0 \; 0]'$ computed from 100 million draws. The left panel, which plots the loss distribution in the loss range [0, 15], suggests that the James-Stein estimator stochastically dominates the MLE. However, close scrutiny of upper quantiles in the right panel, which plots the distribution in the loss range [5, 20], shows otherwise. The CDFs of the two estimators cross at a loss value of approximately

12. Thus, the loss distribution of the James-Stein estimator does not stochastically dominate the one of the MLE.

On the other hand, the loss distribution of the James-Stein positive part estimator does seem to stochastically dominate that of the MLE. In simulations for other parameter values, we have not been able to find any θ for which the loss distribution of the positive part estimator would not seem to stochastically dominate that of the MLE. Based on the simulation evidence, we cannot rule out that the MLE is also SD inadmissible (dominated by the positive part estimator). While we are unable to offer a mathematical proof, the simulation evidence suggests interesting avenues for further research on the Stein phenomenon from the perspective of SD admissibility.

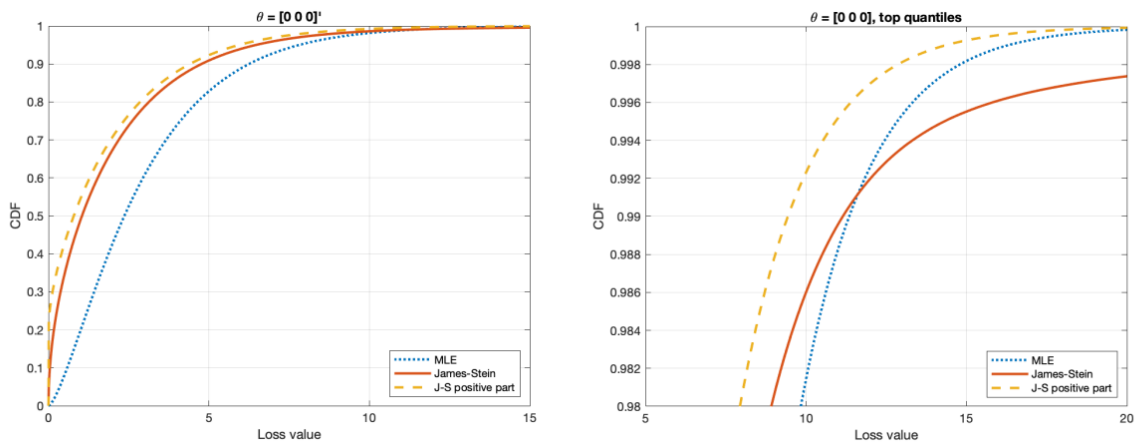

*Figure 1: Comparison of the cumulative loss distributions of the MLE, James-Stein, and James-Stein positive part estimators.*

## 6. Mean and Quantile Decision Criteria

We now turn attention from SD-admissibility to choice of a decision function. Section 2.1 posed three leading criteria that use mean performance to evaluate alternative rules---minimax, minimax-regret, and minimization of Bayes risk. Here we juxtapose these criteria with analogous ones that use quantile performance. Section 6.1 presents the quantile criteria in abstraction. Section 6.2 applies them to selection of a test rule for choice between two treatments.

Decision making using a quantile-utility criterion was proposed in Manski (1988) in a setting without sample data. It was observed there that maximization of expected and quantile utility differ in important respects. Whereas the ranking of actions by expected utility is invariant only to cardinal transformations of the objective function, the ranking by quantile utility is invariant to ordinal transformations. Whereas expected utility conveys risk preferences through the shape of the utility function, quantile utility does so through the specified quantile, with higher values conveying more risk preference. Whereas expected utility is not well-defined when the distribution of utility has unbounded support with fat tails, quantile utility is always well-defined.

There is reason to think that quantiles of welfare distributions matter to decision makers. For example, recent writings on finance have shown explicit concern with low quantiles of earnings distributions, using the term *value-at-risk*. See, for example, Jorion (2006).

### 6.1. Quantile Criteria

Let $\lambda \in (0, 1)$ be a specified quantile. The $\lambda$-quantile minimax and minimax-regret criteria are analogous to criteria (3) and (4) that use risk to measure state-dependent sampling performance. The corresponding quantile criteria are

(16) $\min_{\delta \in \Delta_a} \max_{s \in S} V_\lambda(Q_s\{L[s, \delta(\psi)]\})$,

(17) $\min_{\delta \in \Delta_a} \max_{s \in S} [V_\lambda(Q_s\{L[s, \delta(\psi)]\}) - \min_{d \in D} L(s, d)]$.

There are at least two ways that one might define a quantile analog to minimization of Bayes risk. Replacement of risk with λ-quantile loss in criterion (2) yields

(18) $\min_{\delta \in \Delta_a} \int V_\lambda(Q_s\{L[s, \delta(\psi)]\}) d\pi(s)$.

This hybrid criterion measures the sampling performance of rule δ by the λ-quantile of the sampling distribution of loss and evaluates performance of the rule across states by the mean of these λ-quantiles.

An alternative begins not with definition (2) of the standard Bayes criterion but with the equivalent definition

(2′) $\min_{\delta \in \Delta_a} E\{L[s, \delta(\psi)]\}$,

where $E\{L[s, \delta(\psi)]\} = \int R(s, \delta) d\pi(s) = \iint L[s, \delta(\psi)] d\Phi(s, \psi)$ is the Bayes expectation of loss across samples and states. Replacement of the Bayes expectation of loss with its λ-quantile yields

(19) $\min_{\delta \in \Delta_a} V_\lambda\{L[s, \delta(\psi)]\}$.

Although the mean-based criteria (2) and (2′) are equivalent, the quantile-based criteria (16) and (17) generally differ from one another.

It is well-known that minimization of Bayes risk is also equivalent to solution of the collection of conditional Bayes decision criteria

(2″)   $\min_{d \in D} \int L(s, d) d\Phi(s|\psi), \qquad \psi \in \Psi.$

See, for example, Berger (1985, pp.159-160). That is, minimization of Bayes risk is equivalent to minimization of the posterior expected value of Bayes loss at every point in the sample space. This result, which follows from Fubini's Theorem, does not hold for quantile-based criteria. The quantile analog of (2″) would be to minimize the posterior $\lambda$-quantile of Bayes loss at every point in the sample space. This posterior quantile criterion generally differs from both (18) and (19). We do not know whether it has an interpretation from the perspective of ex ante statistical decision theory. Hence, we do not consider it further.

6.2. Criteria for Selection of a Test Rule

The mean and quantile-based decision criteria of Sections 2.1 and 6.1 offer a menu of procedures for choice of a statistical decision function. To illustrate their application, we continue the analysis of Section 3.2 and consider choice between two treatments, focusing on test rules. Following common practice, we skip the step of determining admissibility and use the criteria to choose among all feasible test rules, not just those that are admissible.

The state-dependent expected and quantile welfare of test rules have simple expressions, namely

(20)   $W(\delta, P_s, Q_s) = \min(\alpha_s, \beta_s) \cdot \rho_s(\delta) + \max(\alpha_s, \beta_s) \cdot [1 - \rho_s(\delta)],$

(21)   $V_\lambda(\delta, P_s, Q_s) = \min(\alpha_s, \beta_s) \cdot 1[\rho_s(\delta) \geq \lambda] + \max(\alpha_s, \beta_s) \cdot 1[\rho_s(\delta) < \lambda].$

Observe that mean and quantile sampling performance are both monotonically decreasing in the error probability, falling from $\max(\alpha_s, \beta_s)$ to $\min(\alpha_s, \beta_s)$ as $\rho_s(\delta)$ increases from 0 to 1. However, they differ in the pattern of decrease. Whereas mean performance varies linearly with the error probability, quantile performance is a step function. This difference in the pattern of decrease implies differences between decision criteria based on mean and quantile performance, described below.

6.2.1. Mean and Quantile Maximin Criteria

Let $\Delta$ denote the space of all test rules. The maximin criteria based on mean and $\lambda$-quantile sampling performance are

(22) $\quad \max_{\delta \in \Delta} \min_{s \in S} \; \min(\alpha_s, \beta_s) \cdot \rho_s(\delta) + \max(\alpha_s, \beta_s) \cdot [1 - \rho_s(\delta)]$,

(23) $\quad \max_{\delta \in \Delta} \min_{s \in S} \; \min(\alpha_s, \beta_s) \cdot 1[\rho_s(\delta) \geq \lambda] + \max(\alpha_s, \beta_s) \cdot 1[\rho_s(\delta) < \lambda]$.

Research on treatment choice evaluating mean sampling performance has dismissed as unpalatable maximin treatment choice based on mean sampling performance. The reason is that criterion (20) is typically solved by a data-invariant rule. Savage (1951) mentions this in passing. Manski (2004) proves it in the special case in which one treatment is a status quo option whose mean treatment response is known.

The reasoning is simple. Let $\alpha_L \equiv \min_{s \in S} \alpha_s$, $\beta_L \equiv \min_{s \in S} \beta_s$, and suppose that $\alpha_L > \beta_L$. There are two data-invariant rules. One always chooses treatment a, yielding error probabilities $\rho_s(\delta) = 0$ in states where $\alpha_s \geq \beta_s$ and $\rho_s(\delta) = 1$ when $\alpha_s < \beta_s$. Minimum expected welfare for this rule is $\alpha_L$. The other always chooses treatment b, yielding $\rho_s(\delta) = 0$ when $\alpha_s \leq \beta_s$ and $\rho_s(\delta) = 1$ when $\alpha_s > \beta_s$. Minimum expected welfare for this rule is $\beta_L$. The maximin criterion prefers the former rule to the latter when $\alpha_L > \beta_L$.

Now consider a data-varying rule, which chooses treatment a for some data realizations and b for others. Suppose, as is typically the case in practice, that $\rho_s(\delta) > 0$ in every state where $\alpha_s \neq \beta_s$. Minimum expected welfare for this rule is less than $\alpha_L$ because there exist states in which $\beta_s < \alpha_L$ and there is positive sampling probability that the rule choose treatment b. Hence, the data-invariant rule that always chooses treatment a uniquely solves the maximin problem.

Maximin choice based on quantile performance does not yield such an extreme result and, hence, may be more palatable. Minimum $\lambda$-quantile welfare with the two data-invariant rules are $\alpha_L$ and $\beta_L$. The minimum $\lambda$-quantile welfare of a data-varying rule is less than $\alpha_L$ if $\rho_s(\delta) \geq \lambda$ in some state where $\beta_s < \alpha_L$. However, the minimum $\lambda$-quantile welfare of such a rule is greater than $\alpha_L$ if $\rho_s(\delta) < \lambda$ in all states where $\beta_s < \alpha_L$ and in some state where $\beta_s > \alpha_L$. Thus, a data-varying rule solves the maximin problem if its error probabilities are positive but not too large.

6.2.2. Mean and Quantile Minimax-Regret Criteria

The minimax-regret criteria based on mean and $\lambda$-quantile sampling performance are

$$(24) \quad \min_{\delta \in \Delta} \max_{s \in S} |\beta_s - \alpha_s| \cdot \rho_s(\delta),$$

$$(24) \quad \min_{\delta \in \Delta} \max_{s \in S} |\beta_s - \alpha_s| \cdot 1[\rho_s(\delta) \geq \lambda].$$

Criterion (24) has been studied by Manski (2004, 2005), Stoye (2009, 2012), Tetenov (2012), and Manski and Tetenov (2016). Criterion (25), which is new, differs because it multiplies the state-dependent loss $|\beta_s - \alpha_s|$ by a step function of $\rho_s(\delta)$ rather than by $\rho_s(\delta)$ itself.

The difference is consequential. The minimax value of mean regret is generically positive. It is zero only in degenerate settings where there exists a rule with $\rho_s(\delta) = 0$ in all states of nature. On the other hand, minimax $\lambda$-quantile regret is zero in some settings with positive error probabilities. Maximum $\lambda$-quantile regret is zero if $\rho_s(\delta) < \lambda$ in all states.

First observe that a rule with $\rho_s(\delta) < \lambda$ in all states trivially exists when $\lambda > \frac{1}{2}$. While one ordinarily thinks of $\psi$ as data that are informative about treatment response, statistical decision theory also encompasses study of STRs that make treatment choice vary with uninformative data. That is, $\delta$ may make the treatment allocation depend on data generated by a randomizing device. Suppose in particular that $\Psi = \{0, 1\}$, $Q_s(\psi = 0) = Q_s(\psi = 1) = \frac{1}{2}$ for all $s \in S$, and $\delta$ is the rule that lets $\Psi_{\delta a} = \{0\}$ and $\Psi_{\delta b} = \{1\}$. The error probabilities for this test rule are $\rho_s(\delta) = \frac{1}{2}$ for all $s \in S$. Hence, the $\lambda$-quantile maximum regret of rule $\delta$ is zero for all $\lambda > \frac{1}{2}$.

To the best of our knowledge, there exists no similarly obvious way to form a rule with zero $\lambda$-quantile maximum regret when $\lambda \leq \frac{1}{2}$. In this domain, achievement of zero maximum regret becomes a more stringent condition as $\lambda$ decreases. It appears infeasible to perform an elementary general analysis, but we can make progress by examining particular contexts.

Proposition 8 demonstrates that test rules with zero maximum regret exist if $S$ is a metric space with positive distance between the sets $S_a$ and $S_b$ (for example, if $S$ is finite), and the data enable sufficiently precise estimation of the true state. In contrast, Proposition 9 shows that for $\lambda < \frac{1}{2}$, no such test rule exists if the set $S$ is connected and other regularity conditions hold. In combination, the two propositions show that zero $\lambda$-quantile maximum regret is neither an empty concept nor ubiquitous. It is attainable by a test rule in some settings but not in others.

*Proposition 8:* Let S be a subset of a metric space $(\Theta, d)$ with distance $d(\cdot, \cdot)$. Let

(26)    $\varepsilon = \frac{1}{2} \cdot \min_{s \in S_a, \, s' \in S_b} d(s, s') > 0$.

Suppose that an estimator $š(\cdot): \Psi \to \Theta$ is $\varepsilon$-far from the true state s with probability below $\lambda$:

(27)    $Q_s[d(š(\psi), s) \geq \varepsilon] < \lambda$ for every $s \in S$.

Then the minimum-distance test rule

(28)    $\delta_{md}(\psi) \equiv 1[\min_{s \in S_b} d(š(\psi), s) < \min_{s \in S_a} d(š(\psi), s)]$

has zero $\lambda$-quantile maximum regret.    □

*Proof:* It follows from the definition of $\varepsilon$ that in every state $s \in S_a$,

$d(š(\psi), s) < \varepsilon \implies d(š(\psi), s) < d(š(\psi), s')$ for all $s' \in S_b$.

The same is true for $s \in S_b$ and $s' \in S_a$. Hence, a necessary condition for rule (28) to yield an error when s is the true state is that $d(š(\psi), s) \geq \varepsilon$. Hence,

$\rho_s(\delta_{md}) \leq Q_s[d(š(\psi), s) \geq \varepsilon]$.

It follows from this and (27) that $\rho_s(\delta_{md}) < \lambda$ for every $s \in S$. Hence, the rule has zero maximum regret.

Q. E. D.

*Remark:* A sufficient condition for (26) to hold is that the average treatment effect $\beta_s - \alpha_s$ be uniformly continuous in s and bounded away from zero. If the state space S is finite, condition (27) is satisfied whenever $š(\cdot)$ is a weakly consistent estimator of the true state and the sample size is sufficiently large.

*Proposition 9:* Let S be a connected subset of a metric space $(\Theta, d)$ with distance $d(\cdot, \cdot)$. Let $S_{a>} \equiv \{s \in S: \alpha_s > \beta_s\}$ and $S_{b>} \equiv \{s \in S: \alpha_s < \beta_s\}$. Assume that the closure of the set $S_{a>} \cup S_{b>}$ is S; that is, for any $s \in S$ and any $r > 0$, there exists $s' \in S_{a>} \cup S_{b>}$ such that $d(s, s') < r$. Let the probability $Q_s(\Psi_0)$ be continuous in s for every measurable subset of the sample space $\Psi_0 \subset \Psi$. Then no test rule with zero $\lambda$-quantile maximum regret exists for $\lambda < \frac{1}{2}$. □

*Proof:* Let $\lambda < \frac{1}{2}$ and suppose that test-rule $\delta$ has zero maximum regret. Let $s_a \in S_{a>}$ and $s_b \in S_{b>}$. Then $Q_{s_a}(\Psi_{\delta b}) < \lambda$ and $Q_{s_b}(\Psi_{\delta b}) > 1 - \lambda$.

Given that $Q_s(\Psi_{\delta b})$ is continuous in s and that S is connected, there exists $s^*$ such that $Q_{s^*}(\Psi_{\delta b}) = \frac{1}{2}$. (See Rudin, 1976, Theorem 4.22). Continuity of $Q_s(\Psi_{\delta b})$ in s implies that there exists $r > 0$ such that $d(s, s^*) < r \Rightarrow |Q_s(\Psi_{\delta b}) - Q_{s^*}(\Psi_{\delta b})| = |Q_s(\Psi_{\delta b}) - \frac{1}{2}| < \frac{1}{2} - \lambda$. Given that $S = cl(S_{a>} \cup S_{b>})$, there exists either $s' \in S_{a>}$ or $s' \in S_{b>}$ such that $d(s', s^*) < r$. Hence, $|Q_{s'}(\Psi_{\delta b}) - \frac{1}{2}| < \frac{1}{2} - \lambda$.

If $s' \in S_{a>}$, the condition $Q_{s'}(\Psi_{\delta b}) < \lambda$ implies that $|Q_{s'}(\Psi_{\delta b}) - \frac{1}{2}| = \frac{1}{2} - Q_{s'}(\Psi_{\delta b}) > \frac{1}{2} - \lambda$, which contradicts the conclusion reached above. If $s' \in S_{b>}$, the condition $Q_{s'}(\Psi_{\delta b}) > 1 - \lambda$ implies that $|Q_{s'}(\Psi_{\delta b}) - \frac{1}{2}| = Q_{s'}(\Psi_{\delta b}) - \frac{1}{2} > \frac{1}{2} - \lambda$, which again contradicts the conclusion reached above. Hence, $\delta$ has positive maximum regret.

Q. E. D.

*Remark*: The state space has the required structure if a is a status-quo treatment with known mean outcome $\alpha^* \in (0, 1)$ and b is an innovation with mean outcome known to lie in the interval $(0, 1)$. Then $S = (0, 1)$, with $\alpha_s = \alpha^*$ and $\beta_s = s$ for $s \in S$. It is the case that $S_{a>} = (0, \alpha^*)$, $S_{b>} = (\alpha^*, 1)$, and $cl(S_{a>} \cup S_{b>}) = (0, 1)$. The sampling distribution has the required continuity if, for example, $Q_s$ is Normal(s, k) for some fixed k $> 0$ or if $Q_s$ is Binomial (n, s) for some integer n.